\documentclass[aip,apl,preprint,groupedaddress,english]{revtex4-1}
\usepackage[T1]{fontenc}
\usepackage[latin9]{inputenc}
\usepackage[letterpaper]{geometry}
\usepackage{xcolor}
\usepackage{float}
\usepackage{amstext}
\usepackage{pslatex}
\usepackage{graphicx}
\PassOptionsToPackage{normalem}{ulem}
\usepackage{ulem}

\makeatletter

\providecolor{lyxadded}{rgb}{0,0,1}
\providecolor{lyxdeleted}{rgb}{1,0,0}

\@ifundefined{textcolor}{}
{%
 \definecolor{BLACK}{gray}{0}
 \definecolor{WHITE}{gray}{1}
 \definecolor{RED}{rgb}{1,0,0}
 \definecolor{GREEN}{rgb}{0,1,0}
 \definecolor{BLUE}{rgb}{0,0,1}
 \definecolor{CYAN}{cmyk}{1,0,0,0}
 \definecolor{MAGENTA}{cmyk}{0,1,0,0}
 \definecolor{YELLOW}{cmyk}{0,0,1,0}
 }

\makeatother

\makeatother

\usepackage{babel}

\begin{document}

\title{Band gap engineering in graphene and hexagonal BN antidot lattices:
A first principles study}

\author{Aihua Zhang$^1$, Hao Fatt Teoh$^2$, Zhenxiang Dai$^1$, Yuan Ping Feng $^1$, Chun Zhang $^{1,3}$}
\email{phyzc@nus.edu.sg}

\affiliation{
	$^1$Department of Physics, National University of Singapore, 
		2 Science Drive 3, Singapore 117542\\
	$^2$Engineering science Programme, National University of Singapore, 
		9 Engineering Drive 1, Singapore 117574\\
	$^3$ Department of Chemistry, National University of Singapore,
		3 Science Drive 3, Singapore 117543}


\begin{abstract}
Effects of antidot lattices on electronic structures of graphene and
hexagonal BN (h-BN) are investigated using the first principles method
based on density functional theory. For graphene, we find that when
the antidot lattice is along the zigzag direction, the band gap opening
can be related to the inter-valley scattering, and does not follow
the simple scaling rule previously proposed in literature for the
antidot lattice along the armchair direction. For h-BN, our calculations
show that the antidot lattice results in reducing of band gaps. Coupled
with doping of carbon atoms, the band gap of an h-BN antidot lattice
can be reduced to below 2~eV, which might have implications in light-emitting
devices or photoelectrochemistry. 
\end{abstract}
\maketitle
Graphene and hexagonal BN (h-BN) have the same two-dimensional (2-d)
honeycomb lattice, but drastically different electronic properties:
Graphene is a gapless semimetal due to the symmetry between two sublattices
of the honeycomb structure, while h-BN is a semiconductor with a wide
band gap (around 5.5 eV~\cite{SongCLSJNKKLYA10}) because of the
lack of such symmetry. Recently, band gap engineering of these two
materials has become a subject of intensive research in the context
of nanoscale electronics~\cite{GeimN07,CiSJJWLSWSBLA10}. Many interesting
ways have been proposed and investigated to tune the band gap of graphene
or h-BN based materials, such as further reducing the material's dimensionality
to form nanoribbons~\cite{PhysRevB.54.17954,SonCL06,PhysRevLett.98.206805},
applying external superlattice potential~\cite{TiwariS09}, decorating
the material with different chemical species (for example, hydrogenation)~\cite{EliasNMMBHFBKGN09,SinghY09,BalogJNARBFLBLSBHPHH10},
and the mixture of these two materials~\cite{CiSJJWLSWSBLA10}. Quite
recently, based on a tight-binding approach (TB), T.\ G.\ Pederson
et al.\ suggested that introducing antidot lattice structure into
graphene may be an effective way to turn graphene from semimetal to
semiconductor~\cite{PedersenFPMJP08}. In their study, for graphene
antidot lattices consisting of a triangular array of holes along the
armchair direction (armchair antidot lattices), a simple scaling rule
for band gaps was proposed, $\Delta E\propto N_{\mathrm{removed}}^{1/2}/N_{\mathrm{total}}$,
where $N_{\mathrm{total}}$ is the total number of carbon atoms in
pristine graphene in one supercell, and $N_{\mathrm{removed}}$ is
the number of carbon atoms removed in the same cell from the pristine
graphene to form the corresponding antidot lattice.

In this paper, first, we investigated electronic structures of another
type of graphene antidot lattices that consists of holes along the
zigzag direction (zigzag antidot lattices). Completely different behaviors
of band gaps from those of armchair cases were found. The inter-valley
scattering is identified as a crucial factor in the gap opening of
graphene antidot lattices and the major reason that causes the differences
between zigzag and armchair cases. We then studied effects of antidot
lattices on electronic properties of h-BN that has the same 2-d structure
but different sublattice symmetry. Interestingly, our calculations
suggested that the inter-valley scattering does not have the significant
effects on band gaps of h-BN as it does to graphene cases, and the
antidot lattice always tends to reduce the band gap of h-BN. With
an appropriate band gap, the engineered h-BN antidot lattice might
be promising not only as an electronic material but aslo in light-emitting
applications or photoelectrochemistry.

\begin{figure}
\begin{centering}
\includegraphics{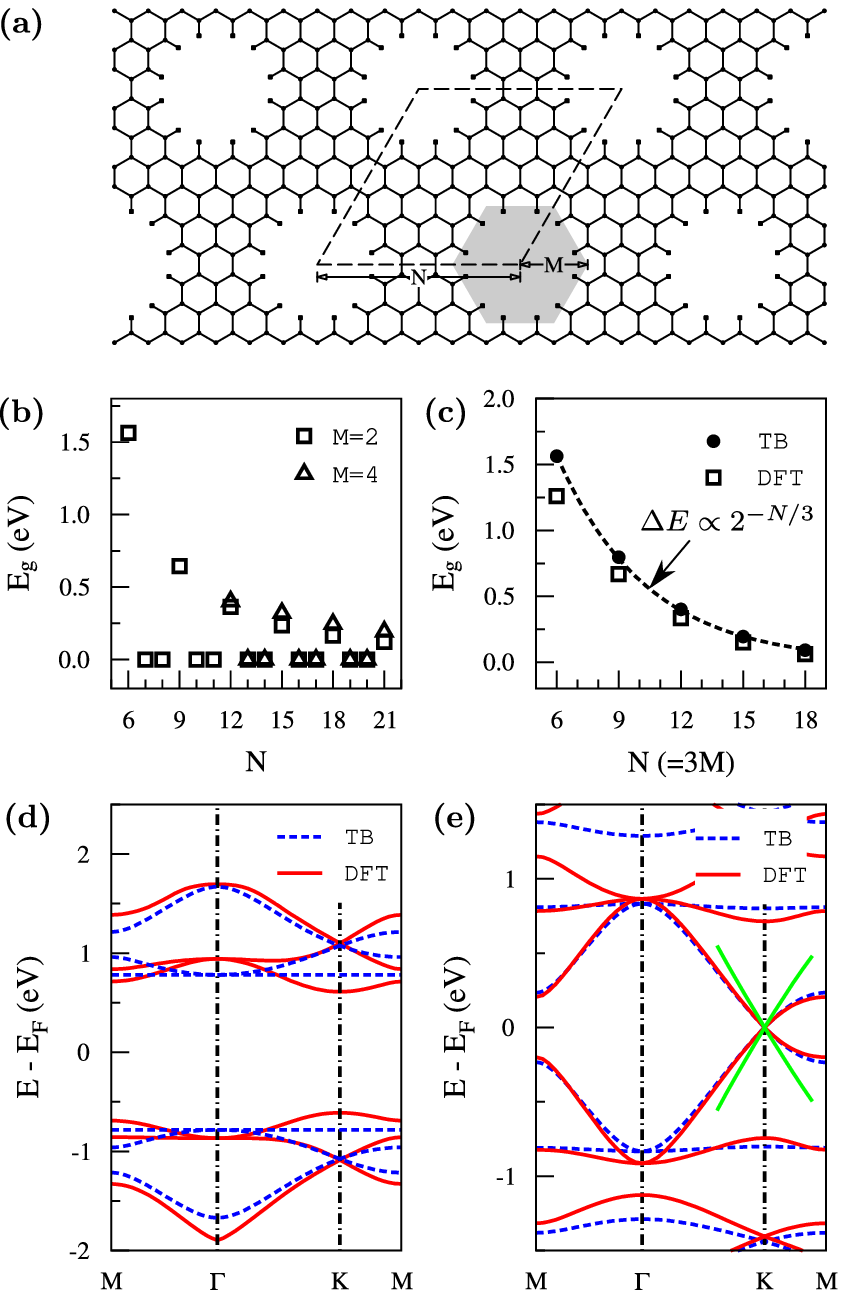} 
\par\end{centering}

\caption{\label{fig:graphene-antidot}(a) A fragment of a triangular graphene
antidot lattice. The unit cell indicated by dashed lines can be characterized
by $(N,\ M)$, where $N$ is the periodicity and $M$ is the side
length of the hexagonal hole. Both $N$ and $M$ are in units of the
graphene lattice constant, $a\approx2.46$~\AA{}. The carbon atoms
at hole edges are terminated by hydrogen atoms (hollow squares). (b)
The variation of band gaps as a function of $N$ with $M=2$ and $M=4$.
(c) The variation of band gaps as a function $N$ with $M=N/3$. The
band gap was calculated using a tight-binding (TB) approach or density
functional theory (DFT). The dashed line shows that TB band gaps can
be well fitted by $\Delta E\propto2^{-N/3}$ with the correlation
coefficient $R^{2}=0.97$. (d) The band structure of a $(6,\ 2)$
antidot lattice. (e) The band structure of a $(7,\ 2)$ antidot lattice.
The graphene band structure around the Dirac point (green lines) is
also shown for comparison.}

\end{figure}

The geometry of graphene antidot lattices studied in this paper is
illustrated in Fig.~\ref{fig:graphene-antidot}(a). The antidots
considered in this study form a triangular lattice and have a hexagonal
shape (as grayed), which is commonly observed in oxygen-etched highly
orientated polycrystalline graphite (HOPG)~\cite{ChangB90,ChuS91}.
We focus on the case that the holes are arranged along the zigzag
direction as shown in the figure. In the following, the graphene antidot
lattice will be referred to as $(N,\ M)$, where $N$ is the supercell
lattice constant and $M$ is the side length of hexagonal holes. Both
$N$ and $M$ are specified in units of graphene lattice constant,
$a\approx2.46$~\AA{}. In such a configuration, numbers of carbon
atoms in two sublattices are equal. Electronic structures are calculated
using both a single-orbital tight-binding (TB) approach~\cite{Wallace}
and density functional theory (DFT). The hopping integral used in
TB calculations is -2.7~eV. DFT calculations were performed using
the \textsc{siesta} code~\cite{SanchezPortalOAS97} with local density
approximation and double-$\zeta$ polarization basis sets. Results
from both TB and DFT calculations agree with each other very well
{[}see Figs.~\ref{fig:graphene-antidot}(c-e){]}. Figure~\ref{fig:graphene-antidot}(b)
shows the dependence of band gaps on the periodicity of holes, $N$,
with hole sizes ($M$) fixed. We observed in this figure that band
gaps open only when $N$ is multiples of three. Figure~\ref{fig:graphene-antidot}(c)
shows the variation of band gaps as a function of $N$ for the case
of $N=3M$. The variation of the band gap in this case can be nicely
fitted with $\Delta E\propto2^{-N/3}$ as shown in the figure, which
is very different from the aforementioned simple scaling rule of armchair
antidot lattices. The corresponding correlation coefficient of fitting
is 0.97. A comparison between band structures of $(6,\ 2)$ and $(7,\ 2)$
configurations is shown in Figs.~\ref{fig:graphene-antidot}(d, e).
The $(6,\ 2)$ configuration, for which $N$ is multiples of three,
has a direct band gap at K. The $(7,\ 2)$ configuration, for which
$N$ is not multiples of three, preserves the characteristic Dirac
cone structure at K, though the Fermi velocity is reduced if compared
to pristine graphene. Such periodicity dependence suggests that band
gap opening in graphene antidot lattices can be related to the inter-valley
scattering between different Dirac points, since when the periodicity
is multiples of three, two nonequivalent Dirac points in prinstine
graphene coincide with each other due to the Brillouin zone folding
and then the inter-valley scattering becomes manifest. If the holes
are along the armchair direction, the band gap is always open because
in this case, the inter-valley scattering condition is always satisfied. In practice, disorder is inevitable in the fabrication process of graphene antidot lattice. To show effects of disorder, we performed a series of calculations with a 48$\times$48 supercell,
in which different holes with $M$ ranging from 2 to 6 and having
an approximate normal distribution of $\overline{M}=4$ and $\mu=1$
are present. Our calculations show that disorder will introduce localized states in the bandgap, and the gap between extended state band edges is larger than that of the ordered lattice.

\begin{figure}[H]
\begin{centering}
\includegraphics[scale=0.8]{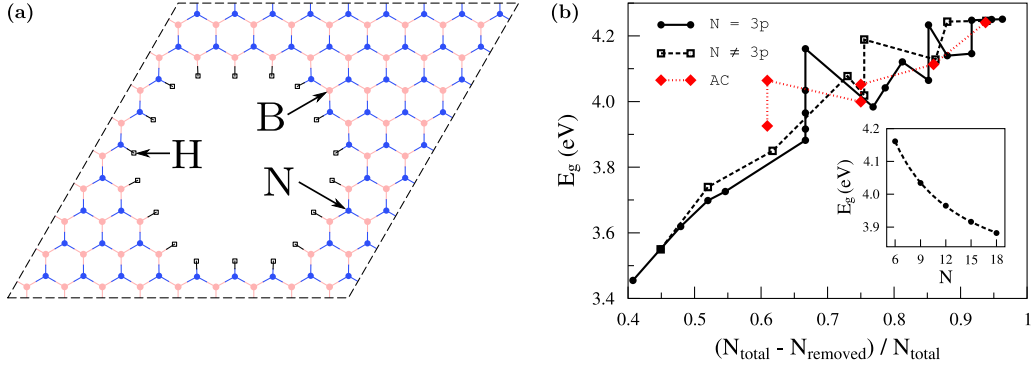} 
\par\end{centering}

\caption{\label{fig:newbnal}(a) The structure of a $(9,\ 3)$ h-BN antidot
lattice. (b) The variation of band gaps as a function of $r\equiv(N_{\text{total}}-N_{\text{removed}})/N_{\text{total}}$
for h-BN antidot lattices along the zigzag ($N=3p$ and $N\neq3p$) and armchair (AC) directions. Generally, the band gap decreases as $r$
decreases and the difference between periodicities of multiples of
three ($N=3p$) and other periodicities ($N\neq3p$) is small, which
might indicate that the inter-valley scattering has a much smaller
effect compared with the smearing effect. In the inset, the variation
of band gaps as a function of $N$ with a fixed $r=\frac{2}{3}$ is
shown and the dashed line shows that the variation can be well fitted
by $\frac{3.9}{N+2.3}+3.7$.}

\end{figure}

We then turn to antidot lattices in h-BN that has the same 2-d atomic
structure as graphene but lacks the sublattice symmetry. The band
gap of pure h-BN predicted by DFT is 4.52~eV, which is about 1~eV
less than the experimental value of multiple-layer
h-BN, 5.5~eV~\cite{SongCLSJNKKLYA10}. The h-BN antidot lattice
is constructed in the same way as the graphene antidot lattice. An
example of $(9,\ 3)$ h-BN antidot lattice is shown in Fig.~\ref{fig:newbnal}(a).
Various kinds of h-BN antidot lattces, $(N,\ M)$, with $N$ ranging
from 6 to 18 and $M$ ranging from 2 to 8, were considered in this
paper. As
shown in Fig.~\ref{fig:newbnal}(b), the band gaps of h-BN antidot
lattices are always smaller than pure h-BN, and unlike the graphene
antidot lattice, there are no significant differences between band
gaps of $N=3p$ cases and those of $N\neq3p$ ($p$ is an integer),
indicating that in h-BN antidot lattices, the inter-valley scattering
does not have significant effects on electronic structures as it does
to graphene antidot lattices. The
band gap variation of h-BN antidot lattices are governed by two trends.
Firstly,
band gaps of h-BN antidot lattices are
overally reduced with the
decreased ratio of $(N_{\text{total}}-N_{\text{removed}})/N_{\text{total}}$. That is to say, with $N$
fixed, a bigger hole size $M$ normally means a smaller band gap,
and with $M$ fixed, a bigger periodicity $N$, usually brings a bigger
band gap. In the Fig.~\ref{fig:newbnal}(b),
the smallest band gap, 3.45~eV, was obtained in h-BN $(18,\ 8)$.
Secondly,
for a fixed ratio, the
band gap scales in an inversely proportional manner with respect to
the size of unit cell, $N$, as shown in the inset of Fig.~\ref{fig:newbnal}(b). 
The band gap scaling behavior is similar if the antidot lattice is along the armchair direction.
The fixed-ratio scaling behaviors are quite different between the
graphene antidot lattice (Fig.~\ref{fig:graphene-antidot}(c)) and
the h-BN antidot lattice, which suggests different mechanisms of band
gap opening for these two materials.

\begin{figure}[H]
\begin{centering}
\includegraphics[scale=1.2]{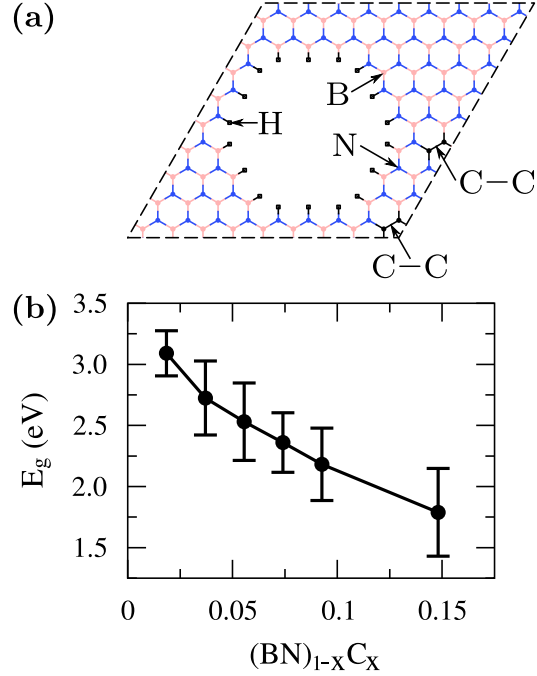} 
\par\end{centering}

\caption{\label{fig:cdoping}(a) An example configuration of two carbon dimers
(indicated by $\mathrm{C-C}$) doped in a $(9,\ 3)$ h-BN antidot
lattice. (b) The variation of band gaps of a carbon-doped $(9,\ 3)$
h-BN antidot lattices as a function of carbon
concentration. }

\end{figure}

The band gap of h-BN antidot lattices may be further reduced by carbon
doping. The carbon doping of h-BN can be done by ion implantation~\cite{PhysRevB.81.245423}.
The essence of the idea is to mitigate the on-site energy difference
between B and N sublattices via random substitution by carbon atoms
on both sublattices. We start with a $(9,\ 3)$ h-BN antidot lattice
which has a band gap of 4.03~eV. We find that it is much more energetically
favorable by more than 1~eV per atom for carbon atoms to form dimers
than to scatter dispersively, which also agrees with experimental
findings~\cite{KawaguchiKN96}, so only doping with carbon dimers
is considered. An example of a configuration with two carbon dimers
is shown in Fig.~\ref{fig:cdoping}(a). For each carbon concentration, which
is defined as the ratio between number of carbon atoms and total number
of atoms except hydrogen, about 20 random doping configurations are
calculated using DFT. The variation of the averaged band gaps with
respect to carbon
concentration and the corresponding sample standard deviations are shown
in Fig.~\ref{fig:cdoping}(b). It can be seen that the band gap decreases
steadily as the concentration of doped carbon dimers increases. A band gap below 2~eV is achievable.
The large range of the band gap adjustable by antidot lattices and
carbon doping, from 4.5~eV to 2~eV, may have applications in semiconducting
electronics, light-emitting devices, and photoelectrochemistry.

In summary, band gap engineering using antidot lattices in graphene
and h-BN is discussed. In graphene with antidots along the zigzag
direction, we show that band gap opening occurs only when the periodicity
is multiples of three, which can be related to the inter-valley scattering.
Due to the lack of sublattice symmetry, the inter-valley scattering
has no significant effects on band gaps of h-BN antidot lattices as
it does to graphene cases. Band gaps of h-BN antidot lattices were
found to approximately scale with $(N_{\text{total}}-N_{\text{removed}})/N_{\text{total}}$.
We further show that the band gap of an h-BN antidot lattice can be
decreased below 2~eV using carbon doping to mitigate the sublattice
on-site energy difference. Both methods have their own merits. Carbon
doping might be adaptive for mass production, while antidot lattices
have the advantage of spatial locality and are suitable for fabricating
of atomic-perfect-interface heterostructures on the same sample. These
two methods combined appropriately might have important implications
in optoelectronics and photoelectrochemistry.

This work was supported by NUS Academic Research Fund 
(Grant Nos: R-144-000-237133 and R-144-000-255-112).

%

\end{document}